\documentclass[journal]{IEEEtran}

\usepackage[cmex10]{amsmath}
\usepackage{cite}
\usepackage{url}
\usepackage{color}

\usepackage{amssymb}
\usepackage{path}
\usepackage{verbatim}
\usepackage{algorithm}
\usepackage{framed}
\usepackage{algpseudocode}

\usepackage{color}
\usepackage{amsmath}
\usepackage{cases}
\usepackage{theorem}
\usepackage{subfig}
\usepackage{subfloat}

\usepackage{graphicx}

\usepackage{pgfplots}
\usepackage{pgfplotstable}
\pgfplotsset{compat=newest}

\newcommand{\bv}{{\bf v}}

\DeclareMathOperator*{\argmax}{\arg\!\max}

\begin{document}

\title{Optimizing the Deployment of Electric Vehicle Charging Stations Using Pervasive Mobility Data}

\author{Mohammad~M.Vazifeh, Hongmou~Zhang, Paolo~Santi and Carlo~Ratti % <-this % stops a space
\thanks{M. Vazifeh and C. Ratti are with Senseable City Lab, Massachusetts Institute of Technology, Cambridge, MA, USA. H. Zhang is with Senseable City Lab and the Department of Urban Studies and Planning, MIT. P. Santi is with Senseable City Lab, MIT, and the Istituto di Informatica e Telematica del CNR, Pisa, Italy. (email correspondence: mvazifeh@mit.edu).}}
%\thanks{Manuscript received...}

\maketitle

\begin{abstract}
With recent advances in battery technology and the resulting decrease in the charging times, public charging stations are becoming a viable option for Electric Vehicle (EV) drivers. Concurrently, wide-spread use of location-tracking devices in mobile phones and wearable devices makes it possible to track individual-level human movements to an unprecedented spatial and temporal grain. Motivated by these developments, we propose a novel methodology to perform data-driven optimization of EV charging stations location. We formulate the problem as a discrete optimization problem on a geographical grid, with the objective of covering the entire demand region while minimizing a measure of drivers' discomfort. Since optimally solving the problem is computationally infeasible, we present computationally efficient, near-optimal solutions based on greedy and genetic algorithms. We then apply the proposed methodology to optimize EV charging stations location in the city of Boston, starting from a massive cellular phone data sets covering 1 million users over 4 months. Results show that genetic algorithm based optimization provides the best solutions in terms of drivers' discomfort and the number of charging stations required, which are both reduced about 10\% as compared to a randomized solution. We further investigate robustness of the proposed data-driven methodology, showing that, building upon well-known regularity of aggregate human mobility patterns, the near-optimal solution computed using single day movements preserves its properties also in later months. 
When collectively considered, the results presented in this paper clearly indicate the potential of data-driven approaches for optimally locating public charging facilities at the urban scale.

\end{abstract}


\IEEEpeerreviewmaketitle

\section{Introduction}

The level of discomfort drivers experiences is, along side high costs, one of the factors contributing to the slow adoption of plug-in all-electric vehicles (PEVs) \cite{Chan1993,Wirasingha2008}. Driver discomfort depends on a number of factors ranging from the limitations imposed by the short vehicle range to the significant amount of time required to recharge  batteries at the charging stations. Current technological advances have increased mileage and reduced the time required for a single full recharging to around 30 minutes \cite{Foley2010,Dickerman2010,Aggeler2010,Etezadi-Amoli2010}. These trends, coupled with the increased push for environment friendly transportation modes, signal the rapidly growing importance of public charging stations as a viable and alternative option to home-stations for charging EVs. An important question to address in this regard is how a network of public charging stations should be deployed in order to minimize the drivers discomfort in an urban environment.

As discussed in detail in the next section, there has been a number of studies aimed at introducing optimization frameworks for the deployment of charging stations. These studies use traditional methodologies developed in the field of facility locations optimization, which are based on computer simulations of the traffic flows and travel surveys \cite{Chen2013,Li,Frade2012,Liu2012,Lam2013,XuK.;YiP.;Kandukuri2013,Hess2012,Wang2013,Xi2013}. Both computer simulations and travel surveys are known to have major limitations in terms of scalability, accuracy, and/or cost. This explains the high interest in the possibility of accessing and analyzing real anonymized individual trajectories to feed the optimization process. This is possible nowadays given the availability of massive data sets of digital traces recording individual-level human movements, such as those provided by cell phone records. These and other data sets collected by urban sensing devices have been used in a number of recent papers \cite{Lee2010, Miluzzo2010, Kwapisz2011, Ni2006, Boulos2011, Wang2011,Boulos2011,Calabrese2010}. However, to our best knowledge data-driven approaches for optimally locating EV charging stations have not been explored so far.

This motivated us to develop the data-driven optimization methodology presented in this paper. The methodology is based on first constructing a model for charging station energy demand that takes into account the mobility patterns of real individual movements recorded in the region surrounding Boston. More specifically, we have used a massive data set of over 1 million cell phone users whose activity has been recorded over a 4 months period. The second step of the methodology consists in formulating a discrete optimization problem to find the optimal configuration for a network of charging stations. To this end, the region of interest is divided into a geographical grid, and an objective function is defined to simultaneously minimize the overall number of charging stations required and the aggregate distance EV drivers need to travel to reach the closest charging station. Since optimally solving the problem at hand is computationally infeasible, we then present computational efficient, near-optimal solutions based on greedy and genetic algorithms. The third and final step is assessing the performance and robustness of the presented approach, which is done by: $i)$ comparing quality of the obtained solutions vs. those provided by a randomized, but locally optimized approach; and $ii)$ showing that the near-optimal solution computed using single day movements preserves its properties also in later months.
%
%Summarizing, the main contributions of this paper are the following: 1) developing a modelling framework for the EV charging demand: 2) formulating a data-driven optimization problem based on facility location theory to leverage the historical data of human movement for efficient transportation planning at urban and regional scale; 2) numerical results on optimization of the configurations based on genetic algorithm and comparing the average distance from charging demand spots to the charging stations for configurations obtained through various heuristic methods; 3) extensive results on the robustness analysis of the obtained patterns with respect to various changes in the model parameters and the time.

The rest of the paper is organized as follows:  in Section \ref{RelWork}, we present an overview of related work. In Section \ref{framework}, we formulate the problem and introduce the data-driven optimization framework. The optimization framework is then described in sections \ref{demand} through \ref{methods}: we describe the mobile phone dataset used in the analysis, and how it has been processed to feed the optimization framework. In Section \ref{results}, we present and discuss the results of the proposed optimization methodology, including a thorough robustness analysis. Finally, Section \ref{conclusions} presents conclusions and discusses future research directions.

\section{Related Work}\label{RelWork}

Facility Location Optimization (FLO) problems have been extensively studied in the field of logistics and transportation planning over the past few decades. With the exception of a single work mentioned below, to our best knowledge the methodologies used for FLO in the context of EV charging station were based on theoretical models, simulations, and/or aggregate transportation data obtained from census tracts or travel surveys. Some of these approaches were based on individual trip trajectory analysis which required theoretical modelling of the behavior of the service seeker agents or travel-surveys. A parking-based assignment introduced in \cite{Chen2013} is based on minimizing EV users' stations access costs while penalizing the unsatisfied demand. The case study relied on 30,000 personal trip records based on household travel surveys in Seattle, US. In \cite{Liu2012}, a multi-objective optimization problem is introduced considering various costs associated with charging stations (including construction, operational costs, charging related costs). The aim of this study was to minimize the costs with consideration of charging demand distribution obtained from aggregate transportation data for Beijing.

In another work, the authors of \cite{Lam2013} proposed an optimization problem from the EV driver's perspective, where the objective is to minimize user discomfort measured as the distance between the charging demand spots and the location of the stations. This study used static population and income-level in Hong Kong as an input to model the spatial distribution of charging demand. In \cite{Hess2012}, EV traffic and drivers' behaviour have been simulated on the real road network of Vienna. Based on a simple linear battery depletion model, an approach is proposed to optimize the location of charging stations for a limited number of EVs with the objective of minimizing the overall travel time.

In \cite{Frade2012}, a study on optimizing the location of EV charging stations for a neighborhood in Lisbon is presented based on estimation of the refuelling demand through the application of a maximal covering model. In this study, the refuelling demand estimation was based on 2001 static census data. In another work, an optimization framework was introduced to deploy charging stations and to assign optimized number of plugs based on real EV taxi-trajectory data, with the objective of minimizing the average time to find a charging station and waiting time before charging \cite{Li}. Using LP-Rounding, the authors provided approximate solutions to the NP-hard optimization problem. To our best knowledge, this is the only study in the field which is based on sensor-based individual travel trajectories data analytics. However, taxi data is known to represent only a small fraction of the actual mobility demand, which can be better characterized by means of cell phone data \cite{Alexander15} as done in this paper.

Other studies have proposed multi-objective optimization problems to locate charging stations. For instance, in \cite{Wang2010a} the authors proposed a model based on set cover with dual objectives of minimizing the overall costs of opening new charging stations and maximizing their overall coverage on Taiwan road network based on static population density data. In \cite{Xi2013}, a linear model is used to estimate EV penetration in various sub-regions to determine the volume of EV flows between the sub-regions based on demographic data. This simulation model is then fed into an optimization problem to obtain the location of charging stations considering an objective function which maximizes the EVs that charge at the new public stations.

The aforementioned works consider various aspects and are important first steps towards constructing a comprehensive optimization framework for efficient planning of EV charging stations. In this work, we further grow the body of knowledge in this important topic by proposing the first optimization framework based on cell-phone data. The importance of relying on cell phone records lies on the fact that a recent study has shown their very good accuracy in modelling individual trip origin/destination flows in urban settings \cite{Alexander15}. We thus believe that cell phone data fed optimization will become a prominent approach for many urban facility location problems, of which EV charging station is a first example.

\section{Introducing the Framework}\label{framework}

\begin{figure}
\begin{center}
\includegraphics[width=0.40\textwidth]{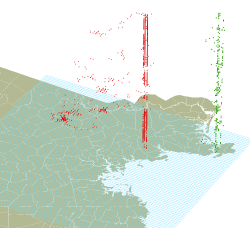}
\caption{The sampled individual location traces across space shown for two different users. The vertical dimension represents time (an entire week in July 2009) in increasing order starting from the plane. Relatively more dispersed mobility of the red vs. the green user is clearly visible.}
\label{fig_schem}
\end{center}
\end{figure}

Our data-driven optimization framework relies on analyzing the movement patterns of individuals obtained through cell-phone data over the span of 4 months. Starting from that, we formulate a discrete optimization problem equivalent to the well-known set cover problem \cite{Cormen09} by dividing the region of interest into a geographical grid. The objective is to minimize the total aggregate distance travelled by drivers on their route from the end of their intended trip to the closest available charging station. Since the set cover problem is NP-hard \cite{Cormen09}, we use Chvatal's greedy \cite{Chvatal1979} and a Genetic Algorithm (GA) \cite{Beasley1996} approach to approximate near-optimal EV charging station locations. 

The overall optimization framework can be divided into the following stages:
\\
\\
A) \emph{Demand Modelling}: Constructing an energy demand model for EVs based on individual trip trajectories
\\
\\
B) \emph{Formulating the Optimization Model}: Formulating the location optimization problem as a set cover problem
\\
\\
C) \emph{Analyzing the Data}: Processing the raw data to estimate EV energy demand based on a simple assumption for EV adoption rates
\\
\\
D) \emph{Optimization Methods}: Performing a number of optimization methods including Chvatal greedy search and GA meta-heuristic search to find near-optimal locations of EV charging stations.

Each step will be described in a separate section.

\section{Step 1: Demand Modelling}\label{demand}

We assume the urban area is partitioned into a number of non-overlapping square cells $\mathcal{C} =\{C_1,C_2,...,C_N\}$, e.g., corresponding to a square grid partitioning. Similar to space, time is also divided into a number of non-overlapping intervals, corresponding to, e.g., $\Delta t = 30~min$ slots. In the following, we use $i$ as a generic cell index and $t$ as a generic time slot index. Equipped with the above definitions, we now define the following quantity for each spatio-temporal slot $(i,t)$:

\begin{figure}
\begin{center}
\includegraphics[width=7.5cm]{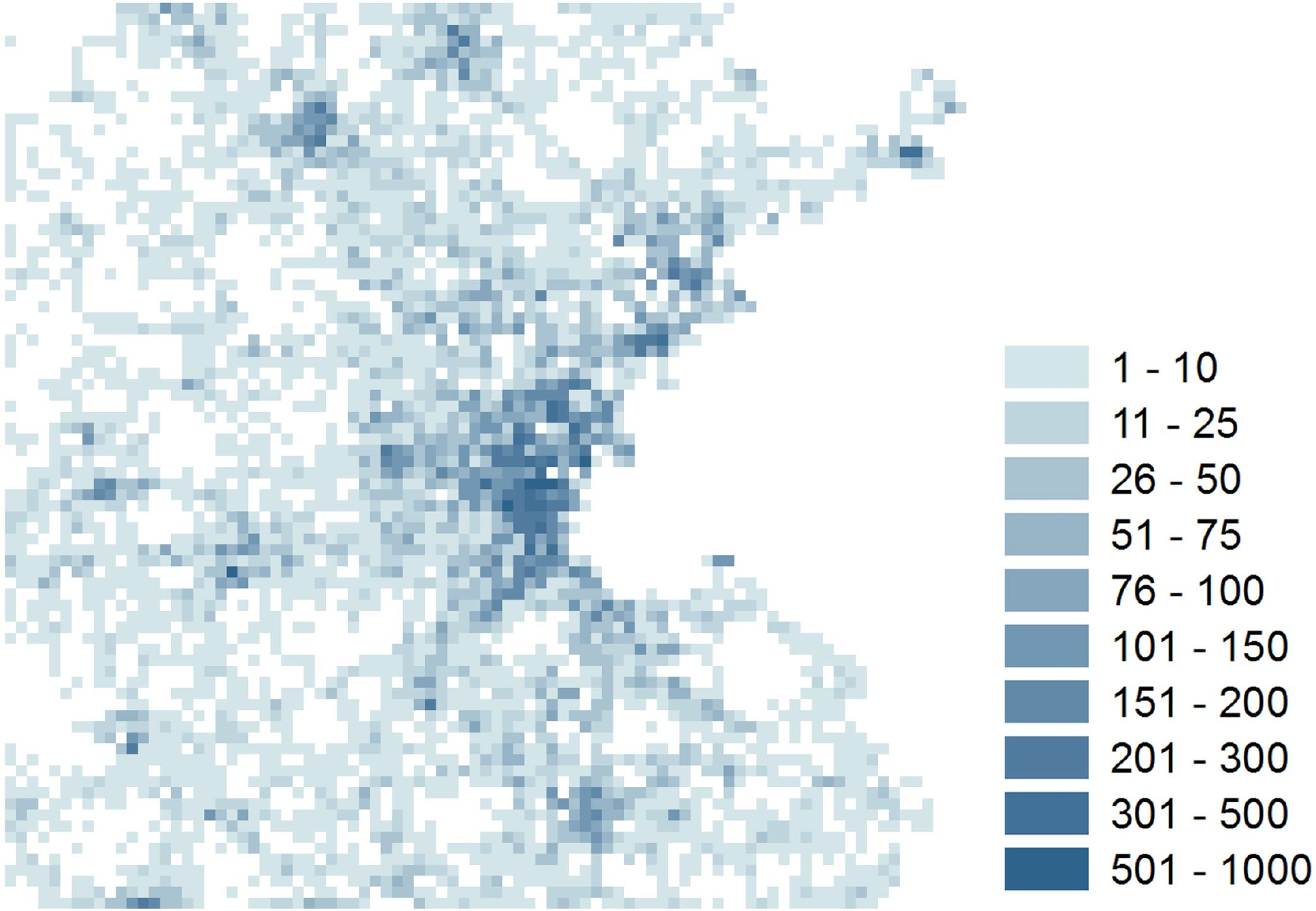}
\caption{The home-location population obtained using the home-detection method described in the text. Higher population density in the Metro Boston area is clearly visible.}
\label{fig_home}
\end{center}
\end{figure}

\begin{gather}
IN_{i,t} = \text{number of trip segments ending}\\ \text{in cell $C_i$ during time interval $[t,t+\Delta t)$}
\end{gather}
We concatenate consecutive trip segments if the stay time between them is less than $\tau_{min}$, where $\tau_{min}$ represents the minimum time interval needed for completely charging an EV, and is set to $30~min$ in the following. The trip segments counted in $IN_{i,t}$ are then those at which ends it is possible to completely charge the EV.

Note that, if we assume that the duration of a time slot $\Delta t$ is $\leqslant \tau_{min}$, all segments contributing to
$IN_{i,t}$ belongs to different vehicles. A fraction $0<\delta\le 1$ of $IN_{i,t}$, i.e., $\delta \cdot IN_{i,t}$, can be defined as the number of candidate EVs arriving and potentially demanding fast refuelling their batteries in $(i,t)$. We construct this refuelling demand model based on the durations of the stay at various cells, and by assuming a linear energy consumption model with the trip length. More specifically, $u_E(t) =  (u_E(0) - c \cdot l) \Theta(u_E(0) - c \cdot l)$ where $u_E$ is $1$ when the EV is fully charged and $0$ when the battery is empty of charge and $\Theta(x)$ is a step function which is zero for $x\leqslant0$ and $1$ when $x>0$. The constant $c$ depends on a number of parameters including road conditions, traffic flow, driver behavior, and vehicle type. For simplicity, we assume that the battery depletion is linear on average with distance travelled, and that the range is $150 km$, which is equivalent to setting $c=1/150 km^{-1}$. In order to have a better estimate of refuelling demand at each cell, at each time slot we only count trip segment end points with trip lengths equal to or larger than $l_{min} \sim 100km$. We obtain the number $IN_{i,t}$ of needed "chargeable" parking spots at each cell according to these considerations. It is important to note that cell phone data does not allow immediately distinguishing between transportation modes. For this reason, we simply assume that the shares of EVs are uniformly distributed in space proportionally to the envisioned penetration ratio $\delta$ of electric vehicles. A better approximation would be to consider a spatial inhomogeneity in the penetration rate as it has been proposed in \cite{Xi2013}, which is left for future work. In the rest of the paper we assume $\delta = 0.25$.

\section{Step 2: Formulating the Optimization Model}

Based on the topology of the cell partitioning, we build a cell network $\mathcal{C}_{N}$ as follows. We
add a node for each cell $C_i \in \mathcal{C}$, and we add a link between $C_i$ and $C_j$ if the two cells are
spatially adjacent to each other. For instance, if $\mathcal{C}$ corresponds to a square grid partitioning,
the node corresponding to cell $C_i$ in $\mathcal{C}_{N}$ is connected to the nodes corresponding to the cells
adjacent to $C_i$ in the North, East, South, and West direction.
We now introduce a parameter $h$, where $h$ is an integer $\geqslant 0$, which models driver discomfort
in terms of how far a driver is willing to drive to reach a charging station. Having set $h$, we
say that a charging station located at $C_i$ {\em covers} a cell $C_j$ if the respective nodes in $\mathcal{C}_{N}$ have
hop-distance at most $h$. In turn, we say that $C_j$ {\em is covered} if there exists at least one cell at
hop-distance at most $h$ from $C_j$ that hosts a charging station.

The tradeoff between drivers discomfort
and the coverage is now clear: if $h = 0$, driver discomfort is minimum, but a charging station
located at $C_i$ covers only the cell itself. With increasing values of $h$, we have a higher driver
discomfort, but coverage of charging stations increases as well, and less charging stations are required to cover the entire city.
Each node in $\mathcal{C}_{N}$ is then weighted as follows. Let $N^{j}_i$ be the set of nodes at hop-distance
$j$ from node $C_i$\footnote{To simplify notation, in the following $C_i$ denotes both a cell and its corresponding node in the cell network $\mathcal{C}_N$.} in $\mathcal{C}_{N}$. The weight $w_i$ of node $C_i$ is computed as

\begin{equation}
w_i = \frac{1}{k} \sum_{s=1,...,k}\sum_{j=0,...,h}\left((j - h) \cdot \sum_{z \in N^{j}_i} \delta \cdot IN_{z,t_s}\right)
\end{equation}

Weight $w_i$ is designed to model the discomfort imposed on drivers by locating charging station
at node $i$. The discomfort is assumed to be linearly proportional to the hop distance to the
charging station, hence it is $0$ for drivers arriving in $C_i$, 1 for drivers arriving in a cell $C_z$ at
hop distance 1 from $C_i$, and so on. So, the total discomfort is obtained by summing up the number
of drivers arriving in cells at a certain hop distance from $C_i$, up to the maximum value of $h$.
Furthermore, this quantity is averaged across the total number of time slots $k$ considered in
the analysis. Note that as we only consider cells in the coverage range of $C_i$ in the computation of $w_i$,
all the weights are negative as $j\leqslant h$.

We are now ready to formulate the charging station optimization problem. The problem of finding the optimal location of charging stations can be formulated as a weighted set-covering problem \cite{Cormen09} in which the universe set, $\mathcal{U}$, and the set of subsets, $\mathcal{S}$, are given as follows
\begin{equation}
\mathcal{U} = \{C_1, C_2, ..., C_N\},
\end{equation}
\begin{equation}
\mathcal{S} = \{s_1, s_2, ..., s_N\},
\end{equation}
where $s_i$ is defined as
\begin{equation}
s_i = \{C_j | C_j \in \bigcup_{z=0}^hN_z^{i}\},
\end{equation}
i.e., the set of all cells within hop-distance $h$ from $C_i$ in $\mathcal{C}_N$.
Each $s_i$ is weighted with $w(s_i)=w_i $, which is a measure of the total discomfort experienced by the drivers in cells covered by $C_i$. The optimization problem is to find a subset $\mathcal{S}_{opt}$ of $\mathcal{S}$ such that all elements of $\mathcal{U}$ are covered, and the sum of the weights in $\mathcal{S}_{opt}$ is minimized. Note that when $h = \infty$, each of the $s_i$'s coincides with the universe set, so the optimal $\mathcal{C}$ is the $s_i$ that has the lowest $w_i$. When $h = 0$, $s_i = \{C_i\}$ and all the weights are zero, therefore optimal $\mathcal{S}_{opt}=\mathcal{S}$ itself, implying that there must be a charging station at every cell. We are then interested in investigating the non-trivial case in which $0<h<\infty$.

The weighted set covering problem described above can be formulated in terms of the following constrained integer linear optimization problem (ILP):

\begin{gather}
\text{Minimize} \sum_{i=1}^{N} (w_i \cdot x_i),
\\
\text{subject to:} \\ p_i = \sum_{j=1,\dots,h}\sum_{z \in N_j^i} x_z \geqslant 1 \;\;\;\; (i = 1,..., N),
\\
x_z \in \{0,1\} \nonumber
\end{gather}

Note that $p_i$ is the number of available stations in the $h$-proximity of $C_i$. In practice, there is a finite capacity on the number of EVs a charging station can handle at a time, i.e., $n_c$. If we require that the capacity is not exceeded, then one need to make sure that the following condition is satisfied for every cell:

\begin{equation}
p_i \geqslant \text{ceiling}\left[\frac{\text{max}(\text{IN}_{i,t})}{n_c}\right] = k_i
\end{equation}

where $\text{max}(\text{IN}_{i,t})=\max_t IN_{i,t}$ is the maximum value of $\text{IN}_{i,t}$ during the observed time interval. Interestingly, this modifies the inequality constraint in a simple way. The difference is that instead of $\geqslant 1$, we have $\geqslant k_i$. Therefore, the problem becomes:

\begin{gather}
\text{Minimize} \sum_{i=1}^{N} (w_i \cdot x_i),
\\
\text{subject to:} \\
 p_i = \sum_{j=1,\dots,h}\sum_{z \in N_j^i} x_z  \geqslant k_i \;\;\;\; (i = 1,..., N),
\\
x_z \in \{0,1\} \nonumber
\end{gather}

In the following, we assume that the capacity of charging stations is large enough to address all energy demand in the coverage range, which is equivalent to setting $k_i = 1$ for all $C_i$s. As discussed in Appendix \ref{capacityCase}, the model can be easily generalized to consider capacitated case for the charging stations. This way, the number of EV charging plugs required in each cell can be optimized to maximize the coverage and to distribute the load evenly between the charging stations.

The set cover problem can be represented also in binary matrix form -- useful for GA implementation -- as follows. An $N\times N$ binary matrix is formed where rows are associated with cells that need to be covered, and columns are associated with stations at various cell locations. Now the $(i,j)$-element of this matrix is either 1 or 0 depending on whether the station at $C_j$ covers $C_i$ or not. This binary matrix, which we call SCP, along with a weights array $W$ uniquely represents the set cover problem:

\begin{eqnarray}
\mathcal{SCP}^h_{ij} =\begin{cases}
1 & \;\;\;\;\; C_j\in\bigcup_{z=0}^h N_z^i\\
\\
0 & \;\;\;\;\;\;\text{otherwise}
\end{cases}~.
\end{eqnarray}
Each configuration of charging stations in this binary representation is a binary vector in which the indices of nonzero elements correspond to indices of the cells which have charging stations in them.

It is also important to note some of the cells in the partitioning have zero energy demand in the observed time period. Therefore, we can reduce the problem size by removing the associated rows/columns.

\section{Step 3: Analyzing the data}\label{data}

\begin{figure*}
\begin{center}
\includegraphics[width=0.8\textwidth]{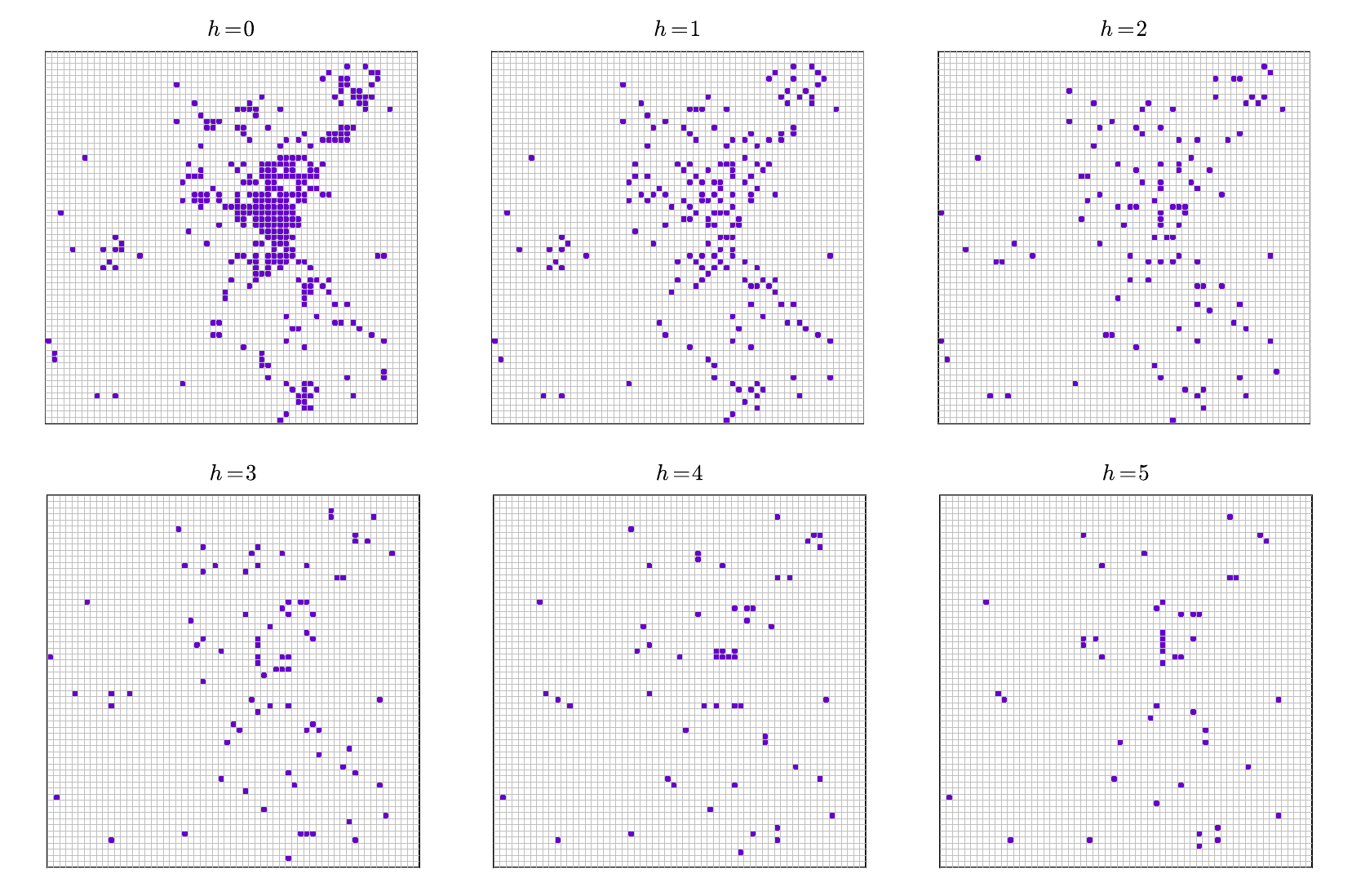}
\caption{Layout of charging stations as computed by the GA for different values of stations coverage $h = 0,1,2,3,4,5$.}
\label{fig1}
\end{center}
\end{figure*}

\subsection{Dataset}

The dataset consists of estimated anonymous location traces from approximately a million users in the Boston Metropolitan Areas, Massachusetts, USA collected by Airsage over the span of 4 months in 2009. The data set contains connection events, including when a text message is sent or received, when a call is placed or received, and when the user connects to the Internet.
The location estimation is generated through triangulation by means of the AirSage's Wireless Signal Extraction Technology. The average inter-event time measured for the whole population is approximately $260~min$. However, the distribution of the inter-event times has an arithmetic average of medians which is around $84~mins$, and the geometric median is around $10.3~min$. This relatively high temporal resolution enables us to detect significant portion of the events associated with location changes of the users. The data set is then suitable to characterize individuals' mobility behavior, which is needed to feed the proposed optimization framework.

\subsection{Processing call data records}
The first data processing step is to identify mobility from the detailed records of cellphone usage. If a continuous sequence of records occurring in the same cell are observed, the user is assumed to stay in that cell from the earliest time in the sequence to the latest. This implies that the records in the middle of the sequence are not relevant for our analysis. Based on this observation, we reduce the list of records from
$$c_{t_0}^{i_0},\ldots,c_{t_l}^{i_0},\ldots, c_{t_k}^{i_p}, \ldots, c_{t_{k+m}}^{i_p},\ldots,c_{t_{n-s}}^{i_q},\ldots, c_{t_n}^{i_q}$$
where $c_{t_k}^{i_p}$ represents a cellphone usage recorded at time point $t_k$ in cell $C_{i_p}$, to a shorter list
$$c_{t_0}^{i_0},c_{t_l}^{i_0},\ldots, c_{t_k}^{i_p}, c_{t_{k+m}}^{i_p},\ldots,c_{t_{n-s}}^{i_q},c_{t_n}^{i_q}$$ obtained removing in-cell records. The list is then transformed into the following list of tuples
$$(i_0, t_0, t_l),\ldots, (i_p, t_k, t_{k+m}),\ldots, (i_q, t_{n-s}, t_n)~.$$
To simplify notation, we relabel this list with continuous subscripts according to their orders in the list, and label the starting and ending times of a location with superscripts `start' and `end'
$$(i_0, t_0^\text{start}, t_0^\text{end}), (i_1,t_1^\text{start}, t_1^\text{end}),\ldots, (i_w, t_w^\text{start}, t_w^\text{end})~.$$

\subsection{Home Detection}
For each user in the data set, we then identify a {\em home cell} amongst the list of visited cells. This step is necessary because we assume that a user does not need to charge her EV at a public charging station if she is at home. Therefore, the arrivals at home location cells will not be counted in $IN_{i,t}$. Fig. \ref{fig_schem} illustrates two sample users' location traces during a week using the vertical axis as time. Each cellphone record is a point on the map based on its location ($(x,y)$ axes) and time of occurrence ($z$ axis).

\begin{figure*}
\begin{center}
\includegraphics[width=0.8\textwidth]{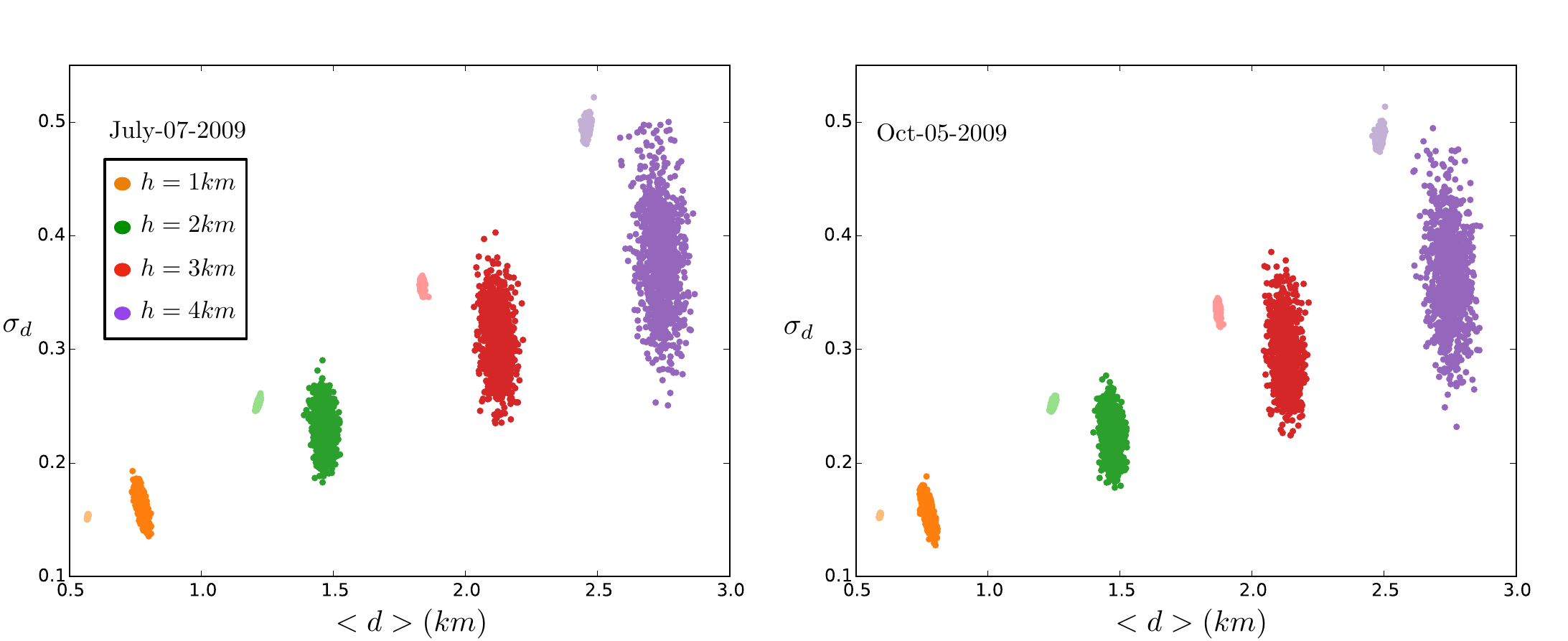}
\caption{The scatter plot of the average driving distance to the closest station, and its variance for a population of the solutions before (larger cluster) and after (smaller cluster) GA. Each point represents a charging station layout.}
\label{fig2}
\end{center}
\end{figure*}

Following the methodology proposed in \cite{Calabrese2010}, the home location $H(u_j)$ of a user $u_j$ is determined by finding the cell where she cumulatively spends most of the time from 8p.m. to 6a.m. in a week, i.e.
$$H(u_j) = C_k,\;\; k = \argmax_p \sum_{s:i_s=i_p} (t_s^\text{end}-t_s^\text{start}) $$
where $t_s^\text{start},t_s^\text{end}$ are only considered if they are between 8p.m. and 6a.m. In Fig. \ref{fig_home} we show the spatial distribution of the identified home locations with a sample of a week's data.

\subsection{Counting the Number of Trips Ending in a Cell}
The process of counting $IN_{i,t}$ is implemented with a linear search. For each user, starting from the beginning of the reduced list of records, we track the total distance the user has traveled from the last charging cell to the current cell through a path along all the intermediate cells without a charging opportunity, i.e., with stays shorter than 30 minutes. For example, if the last charging cell is $C_{i_s}$, since then the user has gone through $C_{i_{s+1}}\ldots,C_{i_{s+k-1}}$, while the current cell is $C_{i_{s+k}}$, the total tracked distance is
 $$ D_{s+k} = \sum_{p=s}^{s+k-1} \mathcal{D}_{i_p,i_{p+1}} $$ where $\mathcal{D}_{i_p,i_{p+1}}$ is the Manhattan distance from $C_{i_p}$ to $C_{i_{p+1}}$.

If during the tracking process, the user stays for longer than 30 minutes at a non-home cell $C_i$ and the total traveled distance is at least $l_{min} = 100 km$, the trip is counted in $IN_{i,t}$ for the respective time period $t$. Then the tracking distance will start from 0 again, until next charging cell is found.

This process is repeated for all the users in the data set, so that the $IN_{i,t}$ values of all cells for all time periods are determined.

\section{Step 4: Optimization methods}\label{methods}

Since optimally solving the set-covering problem is computationally unfeasible, in the following we present two computational efficient methods to produce near-optimal solutions.

\subsection{Greedy search}

The first approach is based on a greedy search \cite{Chvatal1979}, which is used to find a set of column indices that cover all the rows in the binary matrix representation. The algorithm is summarized in the pseudo-code reported in {\bf Algorithm \ref{ps1}}. In this greedy search method, at each optimization step the location of a charging station is chosen by sorting all the available cells according to their average weight per uncovered cell in their coverage range, and a station is placed in a cell corresponding to the lowest value of the average weight per uncovered cell. This process is repeated until all cells are covered. Once the greedy coverage phase is terminate, the algorithm then check whether cells hosting redundant charging stations exist, and, in case they exist, remove them. By eliminating redundant cells, the number of stations will be subsequently minimized. Later on in the paper, we discuss how the objective function can be modified to reduce the number of stations ever further by simultaneously minimizing the number of stations required and the average distance to the closest charging station.

\begin{figure}
\begin{center}
\includegraphics[width=0.45\textwidth]{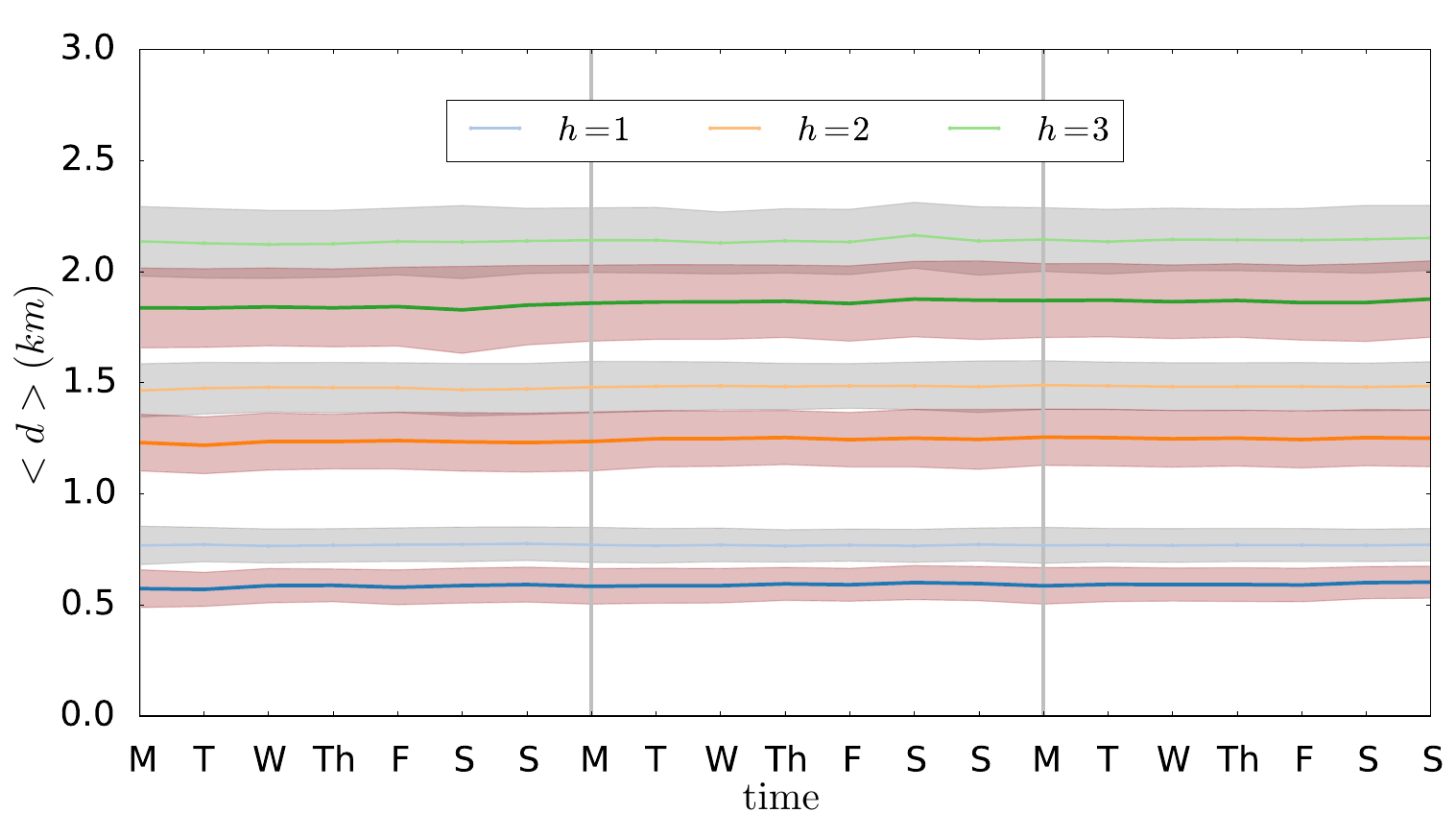}
\caption{The average distance to the charging stations for GA and a randomly selected layout obtained based on a single day in July 2009 computed for the demand on various days in a week selected from the months July, September, and October 2009. The line with darker color correspond to GA and the lighter lines represent a randomly selected solution from the random population of size 1000. The shaded area represents the deviation in distance from the average.}
\label{fig3}
\end{center}
\end{figure}

\subsection{Generating initial GA population}
While providing a provable approximation bound of $O(\log N)$, the solution provided by the greedy search method is known to be inferior to those provided by more sophisticated heuristic/meta-heuristic methods. In this section, we present a Genetic Algorithm (GA) approach. 

The first step is randomly constructing a population of possible layouts by stochastically and uniformly distributing charging stations until all the cells are covered, and then removing the redundant cells that contribute the most to the overall objective function. This way, we can generate a population of locally optimized layouts for charging stations that can be used as the starting point of the GA. Another approach to generate a population of the solutions is a generalization of Chvatal's search method by adding stochasticity to the search scheme. In this probabilistic version of Chvatal's algorithm, deterministic choice of the best possible cell at each stage is replaced by random choice of a cell from a small subset of $k$ elite locations, which are the best $k$ possibilities out of all available cells at each step according to how locations are sorted in  Chvatal's search. The required change in the algorithm is explained in pseudo-code reported in {\bf Algorithm \ref{ps2}}. This way, the solution is no longer deterministic, and each execution of the algorithm results in a different solution. We repeat this process until a population of a desired size is produced. As one increases $k$, the search space becomes more diverse; however, the average quality of the solutions degrades. On the other hand, decreasing $k$ reduces the size of the search space; however, the average quality of the population is higher. In the following, we set $k = 5$, which represents a good compromise between size of the search space and the population quality.

After the initial population is generated, we need to define rules to replace less fit members of the population with new offsprings. Towards this purpose, we closely follow the strategy presented in \cite{Beasley1996}, which uses a binary tournament selection and a fusion crossover operator to evolve the population. The relevant pseudo-code is reported in {\bf Algorithm \ref{BT}, \ref{GA}}. This way the average weight of the population improves and finally the population converges to a high quality set of solutions, which has a significantly lower average distance to the charging stations than that of the best member of the initial population in the most cases -- see next section for details. It is important to note that new binary vectors produced using the fusion crossover operation are not necessarily feasible layouts as they may contain redundant cells or uncovered cells. We need to perform a local search to find a close feasible solution in which all the demand cells are covered and there is no redundancy in the layout. This part is the most computationally expensive part of the algorithm, and it is considered as a necessary step to make sure that the minimum number of stations is used to cover the demand. What we do in this step is very simple: we find all the uncovered cells, and then perform a greedy Chvatal's approximation to cover these remaining cells as we described in the previous section. Then, we remove redundant stations in exactly the same fashion as described in the redundant cell removal stage of {\bf Algorithm \ref{Chvatal}}. To diversify our search space further, we use constant mutation operator which mutates one of the components in the newly generated vector (corresponding to a charging station configuration) in the binary vector representation. The component is chosen among the ones that two parents share with each other. We run the GA by generating $40,000$ children and replacing them with the less fit members of the population. The initial population evolves to a new population after each iteration, till the total number of $40,000$ iterations is reached. 
%\\
%\\
%6) \emph{Robustness Analysis}: Analyzing the robustness of the obtained configurations through the optimization versus demand model parameters

% Insert the algorithm
\begin{algorithm}
\caption{Chvatal greedy approximation}
\label{Chvatal}
\begin{algorithmic}
\State {\em Initialization}
\State $W \leftarrow$ get the weights array
\State $S \leftarrow$ set of all cell indices at which the stations can be located
\State $U \leftarrow$ set of all cell indices which must be covered
\State $\Gamma(s)$ $\leftarrow$ set of cell indices in the coverage range of the element $s$ in $S$, i.e., row indices corresponding to the nonzero elements on the $i$-th column of $\mathcal{SCP}$
\State $\Omega(u)$ $\leftarrow$ set of allowed cell indices which can provide coverage for element $u$ in $U$, i.e., column indices corresponding to the nonzero components on the $i$-th row of $\mathcal{SCP}$
\State $U_c \leftarrow$ set of the so far covered cell indices $\leftarrow$ $\emptyset$
\State $S_c \leftarrow$ set of so far picked columns $\leftarrow$ $\emptyset$
\State {\em Greedy coverage phase}
\While { not $U_c = U$}
\State pick an item $s$ in $S$ that has the minimum weight per its still uncovered rows
\State $S \leftarrow S - \{s\}$
\State $S_c \leftarrow S_c \cup \{s\}$
\State $U_c \leftarrow U_c $ $\cup$ previously uncovered rows covered by $s$
\EndWhile
\State $S_c \leftarrow list(S_c)$ list of picked column indices sorted according to their weights in decreasing order
\State $Q \leftarrow \{S_c\}$
\State {\em Redundant cell removal}
\For {$s$ in $S_c$}
\State $counter \leftarrow 0$
\For {$c$ in $\Gamma(s)$}
\If {$\Omega(c) \bigcap \{Q - \{s\}\} = \emptyset$}
\State $counter \leftarrow counter  + 1$
\EndIf
\EndFor
\If {counter = 0}
\State then $Q \leftarrow Q - \{s\}$
\EndIf
\EndFor
\State \Return Q

\end{algorithmic}\label{ps1}
\end{algorithm}

%\begin{figure*}
%\begin{center}
%\includegraphics[width=0.8\textwidth]{fig1.pdf}
%\caption{Stations location for four different convergence cutoff $h = 0, 1, 2, 3$ for an Inflow matrix averaged over all 96 configurations of the taxi data (taxi?20110215?20110216). In the $h = 0$ case, at every cell with nonzero $\text{IN}_i$, there must be a station. As $h$ increases, the number of stations required to cover cells decreases.}
%\label{fig1}
%\end{center}
%\end{figure*

%\begin{figure*}
%    \centering
%    {\includegraphics[width=0.4\textwidth]{fig5.pdf}}%
%   {{\includegraphics[width=0.4\textwidth]{fig5b.pdf} }}%
%    \caption{Comparison of the distribution of weights for a randomly generated population (green) and a stochastic Chvatal with $k_{max} = 5$ (red) population, both of size 1000.}%
%    \label{dist:compare}%
%\end{figure*}

% Insert the algorithm
\begin{algorithm}
\caption{Stochastic Chvatal}
\label{SChvatal}
\begin{algorithmic}
\State Replace the while loop in the Chvatal greedy approximation with the following
\While { not $U_c = U$}
%\State $k \leftarrow $ pick an integer randomly from $\{1,..., k_{max}\}$
\State $S_k \leftarrow$ pick $k$ item from $S$ that has the least weight per their uncovered rows
\State $s \leftarrow$ choose one item randomly from $S_k$
\State $S \leftarrow S - \{s\}$
\State $S_c \leftarrow S_c + \{s\}$
\State $Uc \leftarrow U_c $ + previously uncovered rows covered by $s$
\EndWhile
\end{algorithmic}\label{ps2}
\end{algorithm}

\section{Results}\label{results}

Fig. \ref{fig1} shows the best configuration obtained for various values of the coverage range $h$, where the $IN_{i,t}$ values are computed from the movement data of a single day in July 2009. The $h=0$ case is the trivial case where there is a charging station at each demand spot, and it just reported to show cells with non-zero energy demand. As one expects, by increasing the value of $h$, a lower number of charging stations is required to cover all the demand. It is interesting to observe that, while the solutions obtained for a relatively larger value of $h$ contains relatively less stations as expected, the computed charging station location sets are {\em not} subsets of those representing previous solutions, indicating the non-trivial combinatorial structure of the problem at hand. 

We can also compute the average distance to a charging station from each demand spot, and compare the initial and final populations of configuration layouts. In Fig. \ref{fig2} {\bf a} we report a scatter plot where each point represents a charging station configuration, and the horizontal and vertical axes represents the average distance and the variance of the distances to the charging stations from the demand spots. Different colors corresponds to different values of $h$. As one can see, for any value of $h$ the final population (smaller clusters) has significantly lower average distance to charging stations compared to initial populations (larger clusters of points). 

An important aspect to investigate is whether the charging station configuration computed using data from a single day is still near-optimal also for the following days. In other words, we would like to assess the robustness of the computed solution in presence of variation of the input data, which is especially important in infrastructure planning processes given the high costs and long lifetime of infrastructure. To this purpose, we have used the near-optimal charging station configuration obtained using a single day in July, and assessed its performance when the $IN_{i,t}$ values are computed from data of various days in a week selected from the months of July, September, and October 2009. As reported in Fig. \ref{fig2} {\bf b}, using the movement data for an equivalent day two months later in October yields very similar results, indicating the robustness of the computed near-optimal configuration. Fig. \ref{fig3} reports the average distance and the variance for the movement data for three weeks chosen from three months of July, September and October, using the same configuration obtained for the single day in July. This plot clearly shows how the achieved improvement in distance remains stable over time. The robustness of the proposed optimization approach is a consequence of the stationarity of the aggregate movement patterns, which have been repeatedly observed in the literature.

\begin{figure}
\begin{center}
\includegraphics[width=0.40\textwidth]{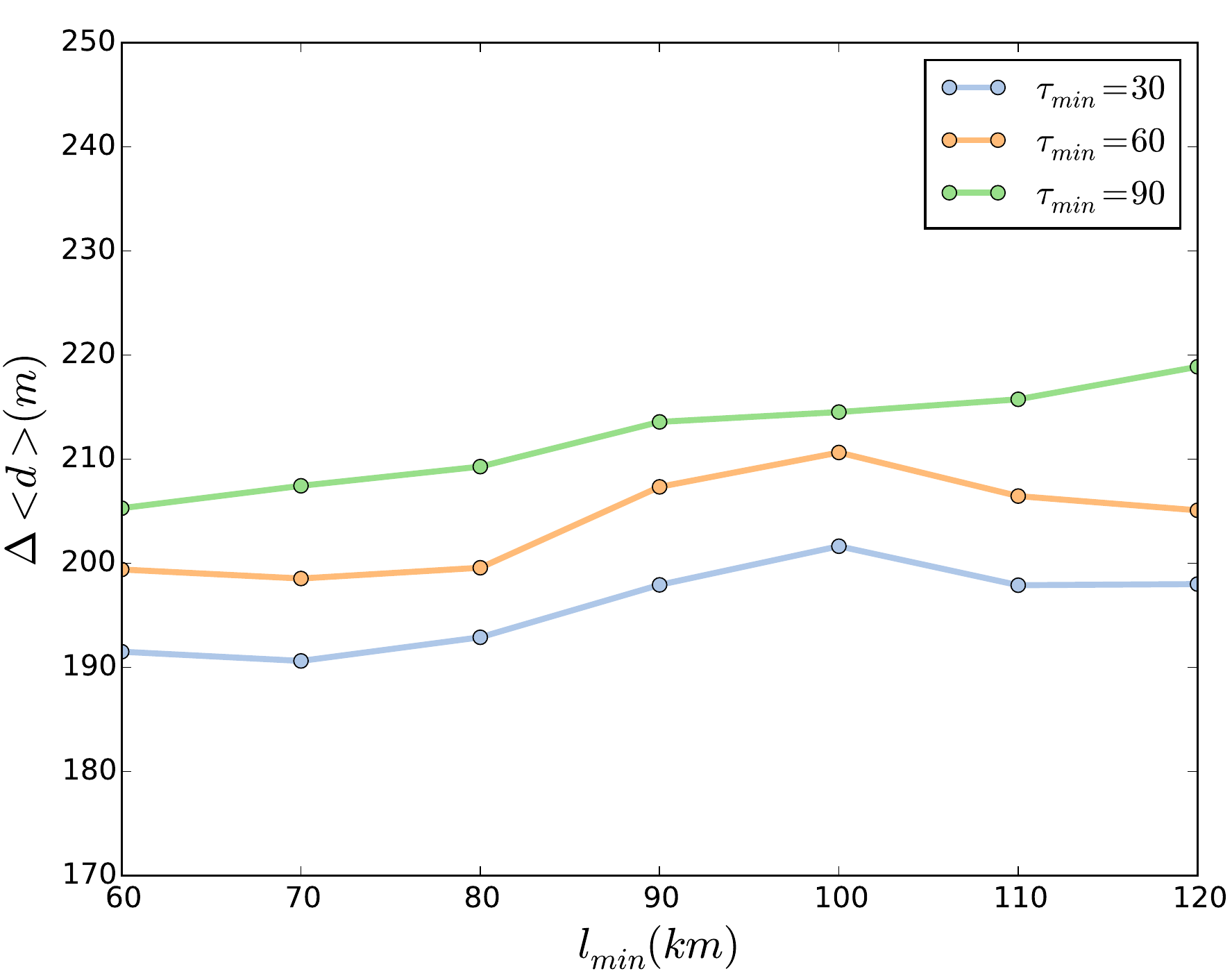}
\caption{The average distance to the charging stations for a layout obtained using GA based on a single day in July 2009, computed for various demand parameters.}
\label{fig5}
\end{center}
\end{figure}

Another way to investigate the robustness of our optimization approach is to tune the parameters of the EV demand to see how sensitive the average driven distance is with respect to a change in these parameters. In Fig. \ref{fig5}, we take the $h=1$ case as a reference, consider the layout configuration obtained for $\tau_{min} = 30~mins$ and $l_{min} =~100km$, and recompute the average distance based on the demand for different values of $\tau_{min}$ and $l_{min}$. Again, the gap in the average distance between a randomly obtained layout and the best layout obtained using GA shows significant robustness in the space of these parameters. This shows that the computed charging station configuration remains stable not only across time, but also in case technological development significantly change charging times and EV ranges. 

Finally, we consider robustness to changes in the EV penetration ratio $\delta$. In this case, it is easy to observe that the structure of the computed solutions is completely independent of the penetration ratio, since $\delta$ is a parameter the uniformly multiplies all weights in the problem formulation. 

\begin{figure}
\begin{center}
\includegraphics[width=0.4\textwidth]{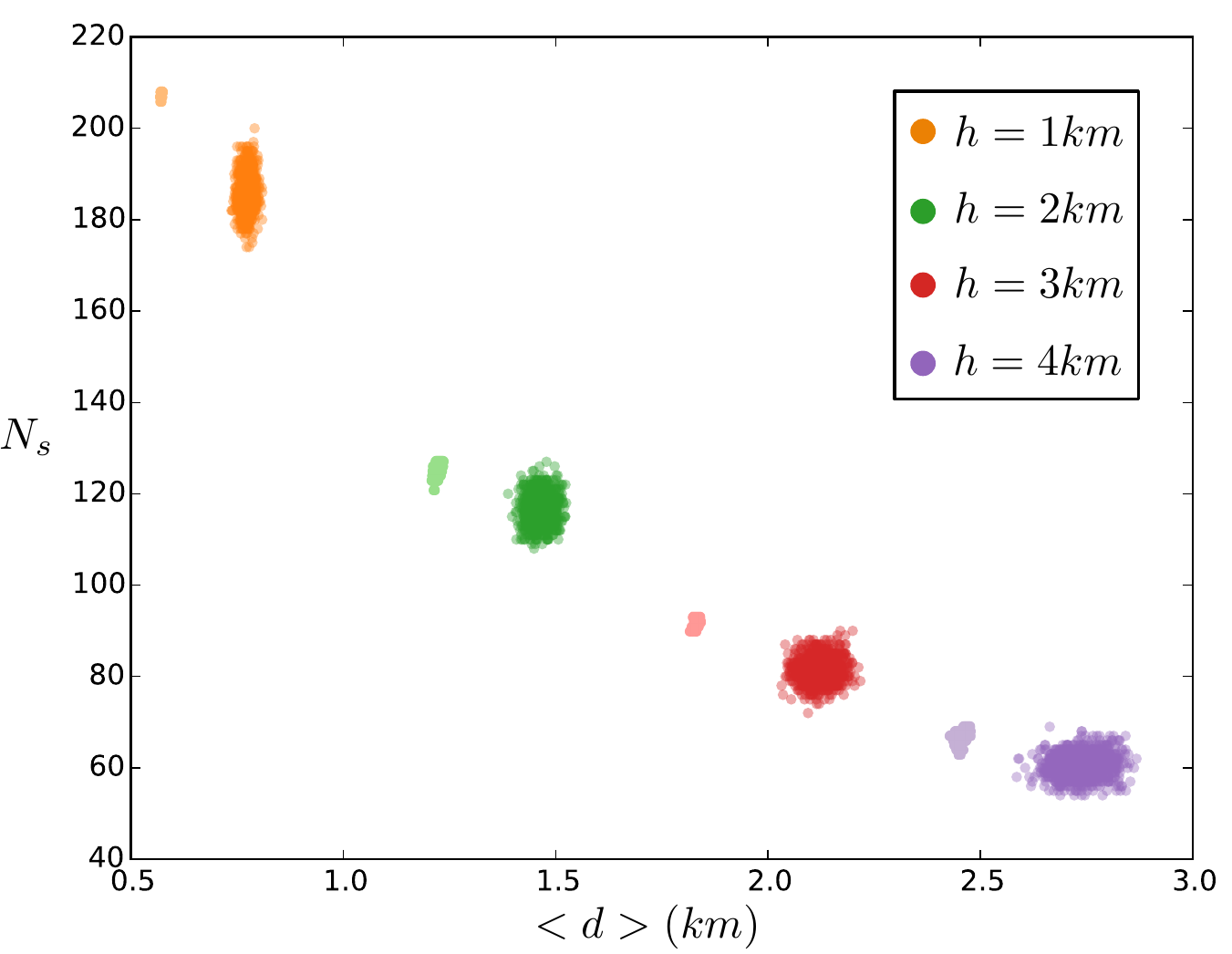}
\caption{The total number of charging stations in each layout versus the average distance to charging stations from the demand spots for set of layouts before (large clusters) and after (small clusters) GA.}
\label{fig4}
\end{center}
\end{figure}

\begin{figure}
\begin{center}
\includegraphics[width=0.4\textwidth]{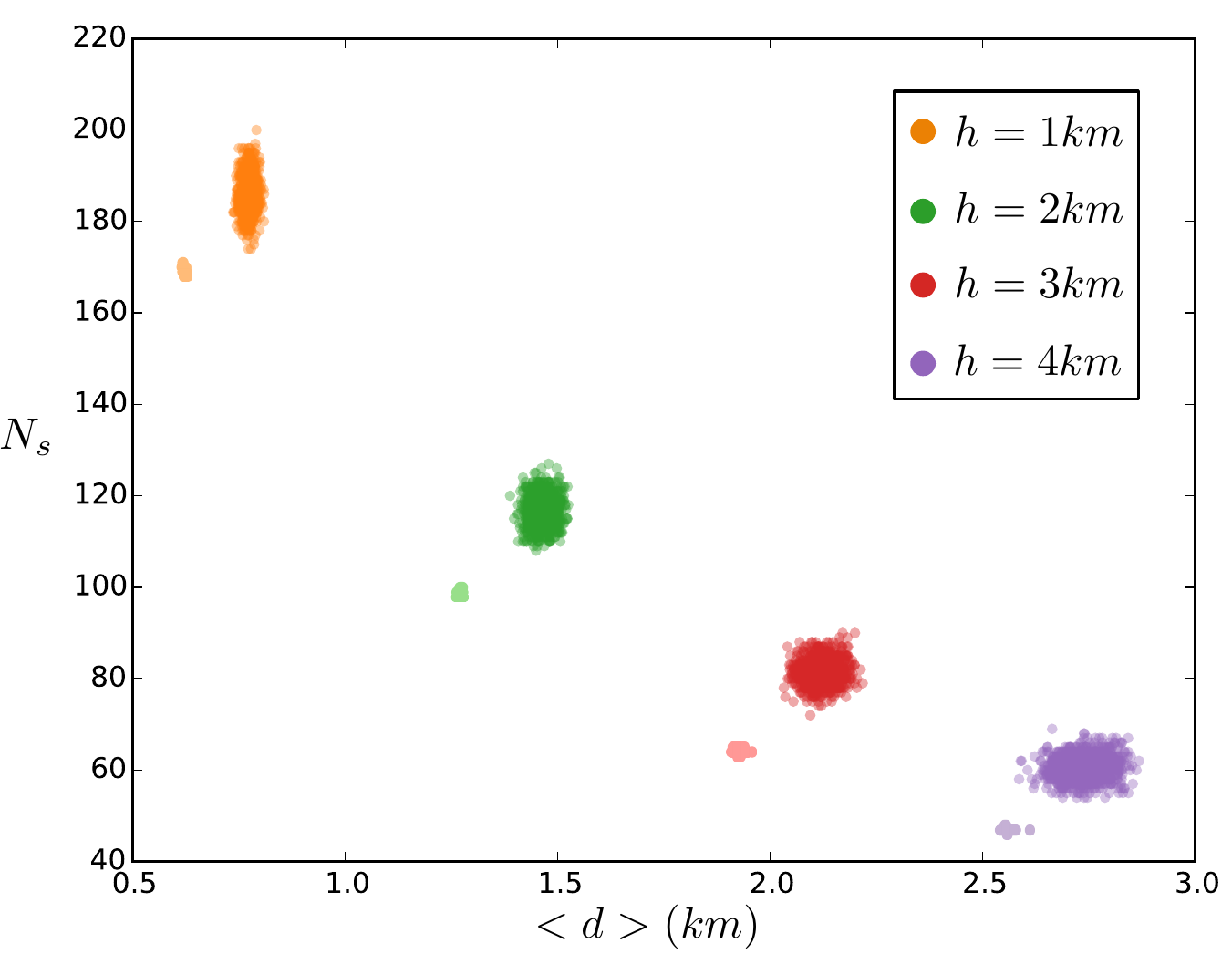}
\caption{The total number of charging stations in each layout versus the average distance to charging stations from the demand spots for set of layouts before (large clusters) and after (small clusters) GA after the objective function is modified to consider the cost of building a new charging station by adding a uniform positive offset, $w_0 = 25 \bar{w}$ to all of the weights, where $\bar{w}$ is the average weight of the cells.}
\label{fig6}
\end{center}
\end{figure}

So far, in the minimal objective function we have not considered the cost associated with building new charging stations. Therefore, it is important to compare the number of stations required in different layouts versus the average distance to make sure the number of charging stations is minimized properly. In Fig. \ref{fig4}, we present a scatter plot of the number of stations versus the average distance to the charging stations for each configuration for charging stations network. This plot shows that layouts with similar numbers of stations can lead to significantly different average distances. The GA population tends to have higher number of stations compared to the average of the initial population. This is consistent with the fact that no penalty term exists in the objective function to penalize addition of new charging stations reflecting the cost of building and operation. To address this issue, we modify the objective function by adding a positive constant $w_0$ to all cell weights. This offset value can in general reflect the cost associated with building a new charging station or other spatial restriction which may exist such as compatibility with the power grid. In the simplest case, we consider a spatially uniform offset and optimize its value to significantly decrease the number of charging stations without too much sacrificing the improvement gained in the average distance. In Fig. \ref{fig6}, we reproduce the result with this modification to the objective function. The number of charging stations is significantly reduced compared to the previous case; however, the improvement in average distance remains still significant in most cases.

% Insert the algorithm
\begin{algorithm}
\caption{Binary Tournament Selection and Fusion Crossover}
\label{BT}
\begin{algorithmic}
\State $N \leftarrow$ Dimension of the binary vectors in the population, which is the same as the number of cells allowed to accommodate charging stations
\State $(\bv_1,\bv_2) \leftarrow$ Randomly select two subset of binary vector members each of size $T=2$, and select the fittest in each subset.
\State $(w_1, w_2) \leftarrow$ Compute the corresponding weights for each of the selected members
\State $p_c \leftarrow \frac{w_2}{w_1+w_2}$
\State $\bv_{new} \leftarrow \bv_1$
\For{$i$ in $\{1,2,...,N\}$}
\If{$\bv_1(i) \neq \bv_2(i)$}
\State $p \leftarrow$ draw uniformly from $[0,1]$
\If{$p > p_c$}
\State $\bv_{new}(i) \leftarrow \bv_2(i)$
\EndIf
\EndIf
\EndFor
\end{algorithmic}
\end{algorithm}
\begin{algorithm}[top]
\caption{Genetic Algorithm}
\label{GA}
\begin{algorithmic}
\State $\mathcal{P} \leftarrow$ generate initial population set of size $M$ using stochastic Chvatal's algorithm
\State $counter \leftarrow 0$
\State $M_c$ number of iterations in the GA
\State $converged \leftarrow$ False
\While{not $converged$}
\State $\bv_{new}$ Generate a new member using Binary Tournament, fusion crossover, and mutation operator
\State $\bv_{new} \leftarrow$ make the new vector feasible using a local Chvatal algorithm.
\If {\textbf{not} $\bv_{new} \in \mathcal{P} $}
\State $\bv_l \leftarrow$ select a member randomly from the population with a weight above $\mathcal{P}$'s average weight
\State $\mathcal{P} \leftarrow \mathcal{P} - \{\bv_l\}$
\State $\mathcal{P} \leftarrow \mathcal{P} + \{\bv_{new}\}$
\State $counter \leftarrow counter + 1$
\EndIf
\If {$counter = M_c$}
\State $converged \leftarrow$ True
\EndIf
\EndWhile
\State $\bv_{best} \leftarrow$ fittest member in $\mathcal{P}$
\State \Return $\bv_{best}$
\end{algorithmic}
\end{algorithm}

\section{Conclusion}\label{conclusions}

Efficient deployment of the network of public charging stations is an important matter which will play a significant role in further increasing the market share of the EVs in the near future.
Motivated by this, we have proposed a modelling and optimization framework to find efficient layout of charging stations to minimize overall EV drivers discomfort, which is reflected in the distance they need to travel to reach the closest charging station starting from their refuelling demand spots. Our minimalistic model can be easily generalized to include more objectives. For instance, spatial variations in the cost of constructing charging stations and the capacity of each station can be considered as an optimization variable without changing the problem structure and the required optimization method. We also showed that the computed near-optimal layout is robust to variation in input data, in EV penetration rate, as well as in EV and charging station technology. This aspect is particularly important given the high costs related to infrastructure deployment, and promote data-driven planning as a viable solution for optimal EV charging station location.

\appendices
\section{Capacitated Case}\label{capacityCase}

Assume that a charging station can handle $l$ EVs at a time. Now if it turns out that, frequently, there are $q > l$ EVs
in a particular cell over a significant time period, then one station is not enough for that cell and that cell needs to
be covered by more than one stations. One way to incorporate this optimization dimension in the context of
the set-covering problem is to duplicate the corresponding $C_i$ in the universe set. In other words, if it turns out that
cell $C_i$ needs to be covered by at least $k_i$ stations, we can redefine the universe set $\mathcal{U}$ by replacing the $C_i$ element in it with $C_i^1,C_i^2,...,C_i^{k_i}$. The weight needs also to be modified accordingly: $\text{IN}_{i,t}\rightarrow \text{IN}_{i,t}/k_i$. With these modifications, the universe set grows, and more subsets are required to cover the whole set. Also, the discomfort contribution of cell $C_i$ decreases for a station located at distance $j$, because its load is now divided between the other $k_i - 1$ available stations\footnote{This holds true under the assumption that drivers have access to real time information about the availability of the stations.}. The interesting point is that the problem is still a set covering problem.

%
%\section{Distance matrix for weight computation}\label{distance}
%Assuming that we have a way to extract numbers associated with the EV traveller users, we can calculate the weights as follows:
%
%\begin{equation}
%w^h_i = \frac{1}{k}\sum_{s=1}^{k}\sum_{j=1}^{N}\mathcal{D}_{ij}^{h} \cdot N^{\text{in}}_{j,t_s}
%\end{equation}
%
%where $w^h_i$ is the weight associated with a charging station with coverage $h$ at cell $C_i$. $\mathcal{D}_{ij}^h$, in the simplest model, is defined in such a way that it is zero when $C_j$ cell is not within the coverage cutoff $h$ assuming that we are using a hard cutoff (which we may want to relax as we discuss later). If $C_j$ is in the cutoff range then $\mathcal{D}_{ij}^h$ is equal to the $ij$ component of the effective distance matrix $\mathcal{D}_{ij}$ .
%
%The effective distance matrix depends on the road infrastructure in a city. If we assume that the grid used to partition the map is in accordance to the city blocks then one simple approximation to the effective distance would be:
%
%\begin{equation}
%\mathcal{D}_{ij} = |X_i - X_j| + |Y_i - Y_j|,
%\end{equation}
%
%in which $X_i$ and $Y_i$ are the $(x,y)$ coordinate of the $C_i$. Therefore, in this simple approximation we have
%
%\begin{eqnarray}
%\mathcal{D}^h_{ij} =\begin{cases}
%|X_i - X_j| + |Y_i - Y_j| & \;\;\;\;\; \mathcal{D}_{ij} < h \\
%\\
%0 & \;\;\;\;\;\;\;\;\;\;\;\;\;\;\;\text{otherwise}
%\end{cases}
%\end{eqnarray}
%
%
%
%

%% you can choose not to have a title for an appendix
%% if you want by leaving the argument blank
%\section{}

\section*{Acknowledgment}

The authors thank ENEL Foundation, Accenture China, American Air Liquide, Emirates Integrated Telecommunications Company (du), Ericsson, Kuwait-MIT Center for Natural Resources and the Environment, Liberty Mutual Institute, Singapore-MIT Alliance for Research and Technology (SMART), Regional Municipality of Wood Buffalo, Volkswagen Electronics Research Lab, and all the members of the MIT Senseable City Lab Consortium for supporting this research.

\bibliographystyle{ieeetr}
\bibliography{EV.bib}
\vspace{-0.5cm}
\begin{IEEEbiography}[{\includegraphics[width=1in,height=1.25in,clip,keepaspectratio]{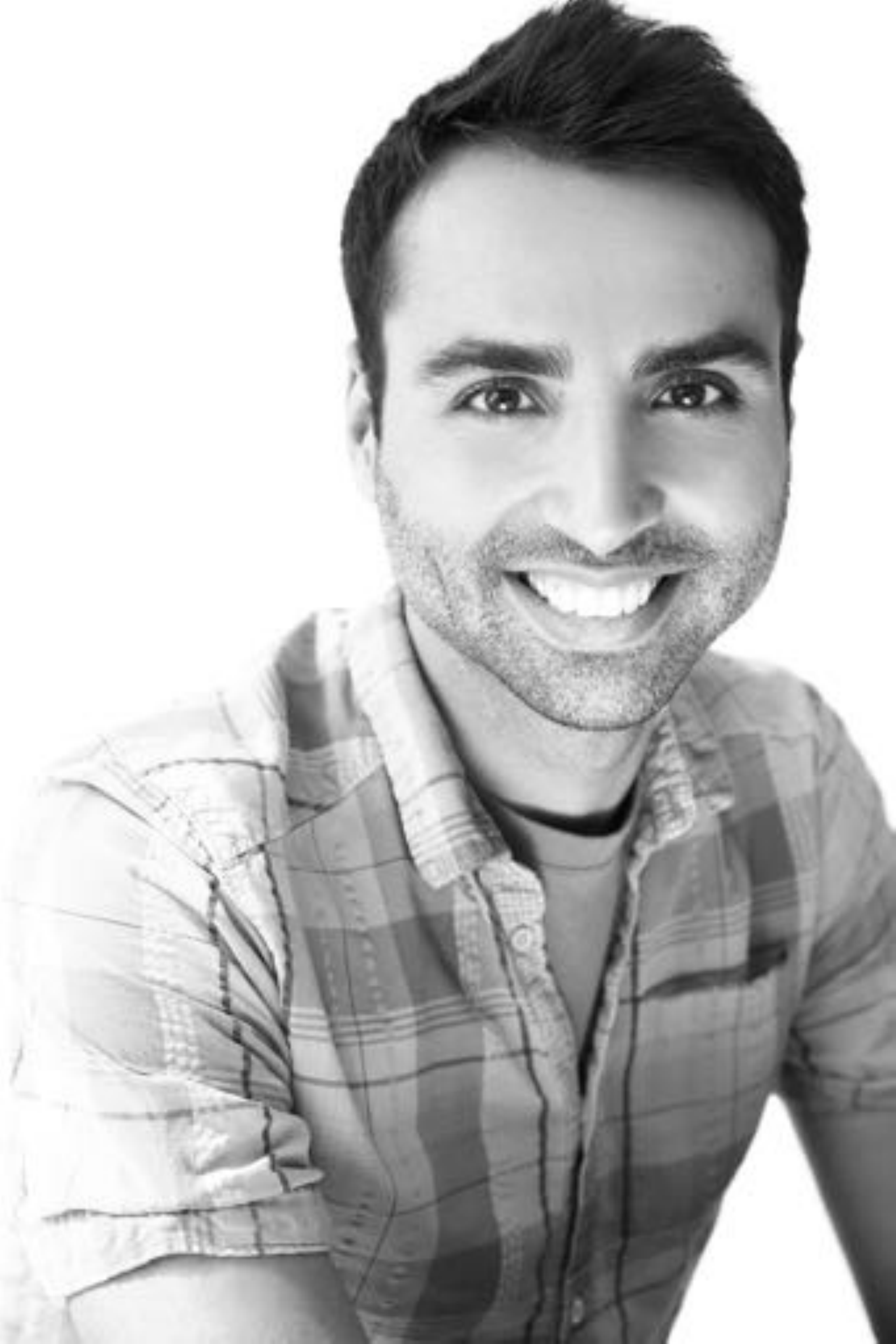}}]{Mohammad Vazifeh}
received the Ph.D. degree in physics from the University of British Columbia (UBC), Vancouver, Canada. He is currently a Postdoctoral Fellow at MIT Senseable City Lab. His background is in theoretical condensed matter physics. Since 2015, his research focus has shifted to machine learning, optimization algorithms design and urban computing. Before joining Senseable City Lab, he worked as a senior algorithm researcher, designing optimization algorithms for the quantum computing hardware with adiabatic quantum annealing architecture. At MIT, he is involved in a number of urban mobility projects including modelling human mobility across urban and regional scale and data-driven optimization in urban infrastructure planning. Currently, his main topic of interest is modelling and predicting human mobility in cities.
\end{IEEEbiography}

\begin{IEEEbiography}[{\includegraphics[width=1in,height=1.25in,clip,keepaspectratio]{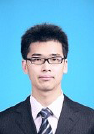}}]{Hongmou Zhang}
received the Master of City Planning degree from the University of Pennsylvania. He is currently a Ph.D. student at the Department of Urban Studies and Planning, Massachusetts Institute of Technology. His current research interests are urban modelling and simulation with massive mobility data, and behavioural modelling in transportation.
\end{IEEEbiography}

%\newpage

\begin{IEEEbiography}[{\includegraphics[width=1in,height=1.25in,clip,keepaspectratio]{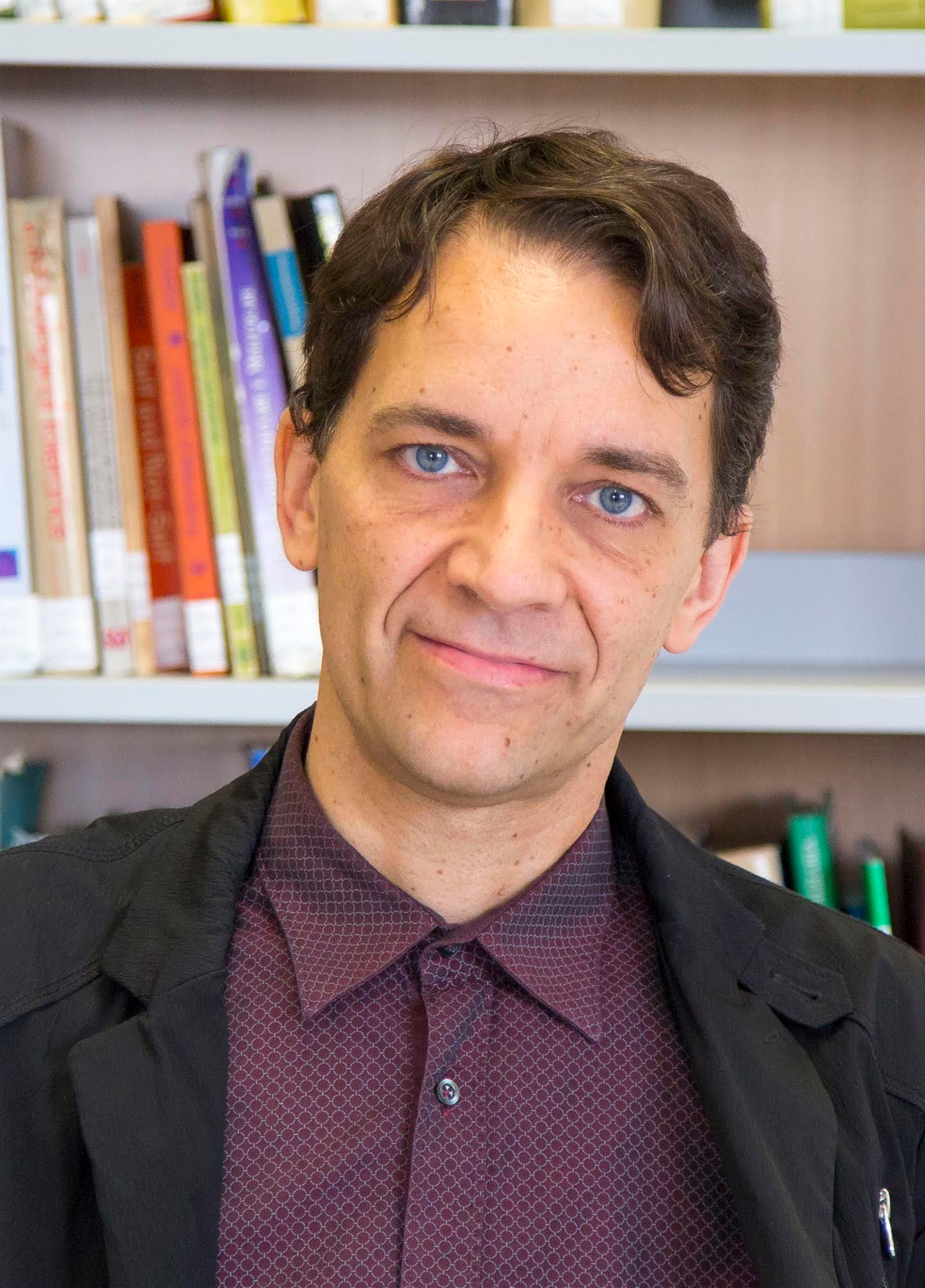}}]{Paolo Santi}
Paolo Santi is a Research Scientist at MIT Senseable City Lab where he leads the MIT-Fraunhofer Ambient Mobility initiative, and he is a Senior Researcher at the Istituto di Informatica e Telematica del CNR. His research interest is in the modeling and analysis of complex systems ranging from wireless multi hop networks to sensor and vehicular networks and, more recently, smart mobility and intelligent transportation systems. In these fields, he has contributed more than 100 scientific papers and two books. Dr Santi has been Guest Editor of the Proceedings of the IEEE, and has served in the Editorial Board of IEEE Transactions on Mobile Computing and IEEE Transactions on Parallel and Distributed Systems. Dr Santi has recently been recognized as Distinguished Scientist by the Association for Computing Machinery.
\end{IEEEbiography}

\begin{IEEEbiography}[{\includegraphics[width=1in,height=1.25in,clip,keepaspectratio]{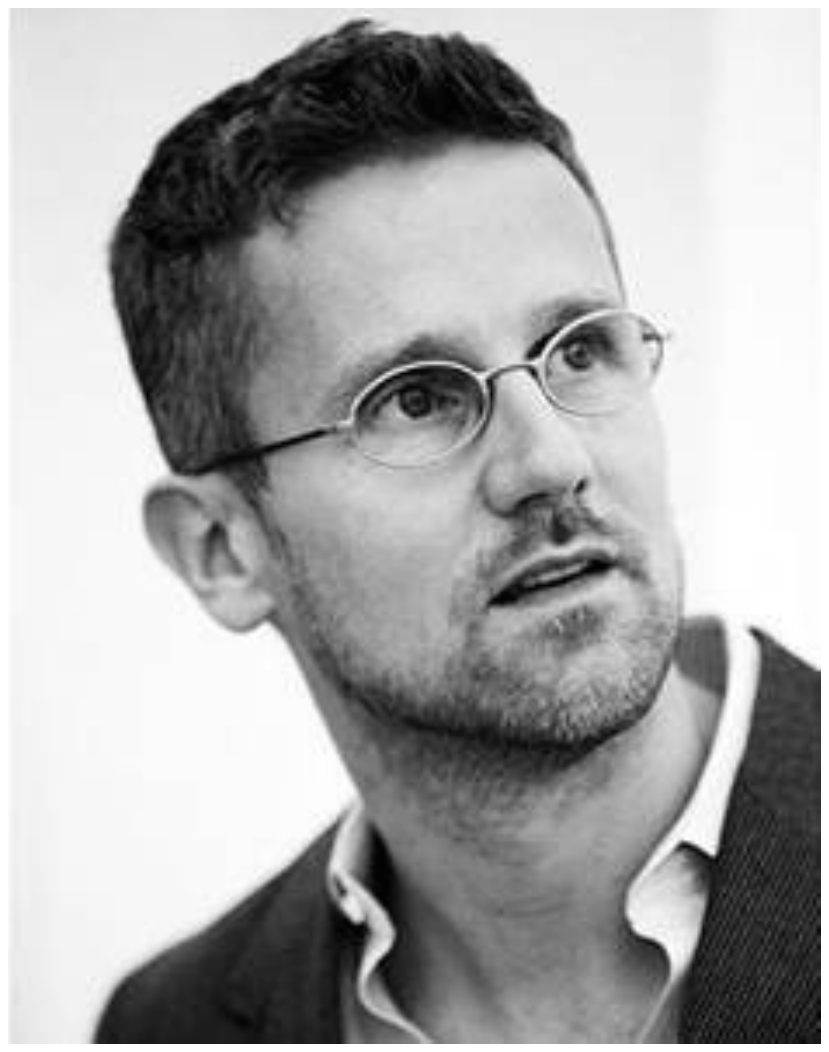}}]{Carlo Ratti}
received the Ph.D. degree from the
University of Cambridge, Cambridge, U.K.
He is currently the Director of the Senseable
City Laboratory, Massachusetts Institute of Technology,
Cambridge, and an Adjunct Professor with
Queensland University of Technology, Brisbane,
Qld., Australia. He is also a founding partner (with
W. Nicolino) and the Director of the architectural
firm Carloratti.
Dr. Ratti is a member of the Ordine degli Ingegneri
di Torino and the Association des Anciens Elèves
de l'\'Ecole Nationale des Ponts et Chauss\'ees.
\end{IEEEbiography}


%\vfill



\end{document}